\pdfoutput=1

\documentclass[11pt,twoside,a4paper,cmspaper,final,collab]{cms-tdr}
\def\svnVersion{267150}\def\cmsCernNo{2013-037}\def\cmsCernDate{\today}\def\cmsMessage{Published in the Journal of High Energy Physics as \href{http://dx.doi.org/10.1007/JHEP08(2011)005}{\doi{10.1007/JHEP08(2011)005.}}}

\begin{document}\cmsNoteHeader{EXO-11-065}

\hyphenation{had-ron-i-za-tion}
\hyphenation{cal-or-i-me-ter}
\hyphenation{de-vices}

\RCS$Revision: 67583 $
\RCS$HeadURL: svn+ssh://svn.cern.ch/reps/tdr2/papers/EXO-11-065/trunk/EXO-11-065.tex $
\RCS$Id: EXO-11-065.tex 67583 2011-07-12 12:08:42Z claudioc $
\newcommand{\met} {\ensuremath{E\!\!\!\!/_T}}
\providecommand{\PQu}{\cPqu\xspace}
\providecommand{\PAQu}{\cPaqu\xspace}
\providecommand{\PQb}{\cPqb\xspace}
\providecommand{\PAQb}{\cPaqb\xspace}
\providecommand{\PAQt}{\cPaqt\xspace}
\providecommand{\PQt}{\cPqt\xspace}
\providecommand{\PAp}{\Pap\xspace}
\providecommand{\PZpr}{\cPZpr\xspace}
\providecommand{\PGm}{\Pgm\xspace}
\providecommand{\PGmp}{\Pgmp\xspace}
\providecommand{\PBd}{\ensuremath{\cmsSymbolFace{B_{d}}}\xspace}
\providecommand{\PABd}{\ensuremath{\cmsSymbolFace{\overline{B}_{d}}}\xspace}
\newcommand{\pp}{\Pp{}\Pp}
\newcommand{\ppbar}{\Pp{}\PAp}
\newcommand{\uubar}{\PQu{}\PAQu}
\newcommand{\tbartbar}{\PAQt{}\PAQt}
\cmsNoteHeader{EXO-11-065} 
\title{Search for Same-Sign Top-Quark Pair Production at $\sqrt{s}~=~7$~TeV and Limits on Flavour Changing Neutral Currents in the Top Sector}

\date{\today}

\abstract{
An inclusive search for same-sign top-quark pair production
in $\pp$ collisions at $\sqrt{s} = 7$ TeV is performed using
a data sample recorded with the CMS detector in 2010, corresponding
to an integrated luminosity of 35\pbinv.
This analysis is motivated by recent studies of
$\ppbar \to \ttbar$
reporting mass-dependent forward-backward asymmetries larger
than expected from the standard model.  These asymmetries could be due
to Flavor Changing Neutral Currents (FCNC) in the top sector induced by
$t$-channel exchange of a
massive neutral vector boson (\PZpr).
Models with such a \PZpr also
predict enhancement of same-sign top-pair production in
$\pp$ or $\ppbar$ collisions.
Limits are set as a function of the \PZpr mass and
its couplings to \PQu and \PQt quarks.
These limits disfavour the FCNC interpretation of the Tevatron results.
}

\hypersetup{%
pdfauthor={CMS Collaboration},%
pdftitle={Search for Same-Sign Top-Quark Pair Production at sqrt(s) = 7 TeV and Limits on Flavour Changing Neutral Currents in the Top Sector},%
pdfsubject={CMS},%
pdfkeywords={CMS}}

\maketitle 

\begin{figure}[htb]
\begin{center}
\includegraphics[width=0.35\linewidth, height=0.25\linewidth]{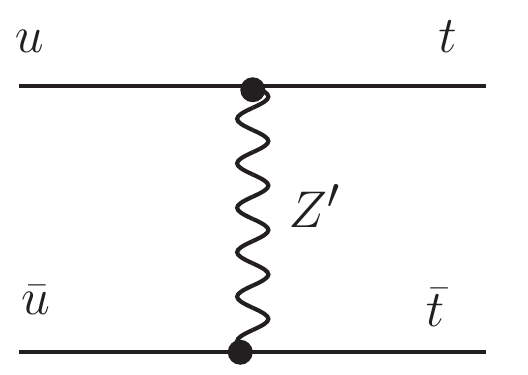}
\caption{ Diagram for $\ttbar$ production induced by
$t$-channel \PZpr exchange, which
can generate a forward-backward asymmetry. \label{fig:ttbar}}
\end{center}
\end{figure}

The Compact Muon Solenoid (CMS)
Collaboration recently published a search for new physics in
events with same-sign isolated dileptons, jets, and \met~\cite{sspaper}.
The results of that search are recast in this
short Letter to set constraints on the production of same-sign top-quark pairs at
the Large Hadron Collider (LHC).
This effort is motivated by the
recent Tevatron measurements of the forward-backward
$\ttbar$ asymmetry ($A_{FB}$)
which deviates from the standard model (SM)
expectations~\cite{d0:fwtop, cdf:fwtop1, cdf:fwtop2}.
Many of the attempts put forth~\cite{th0, Jung:2010yn, th1, th2, th3, berger,
th4, th5, th6, th7, th8, th9, th10, Gresham, Buckley, zoltan,
cdfth, yung, both1, both2, both4} to explain this
asymmetry invoke Flavour
Changing Neutral Currents (FCNC) in the top-quark sector~\cite{th0}
mediated by the $t$-channel
exchange of a new massive \PZpr boson, as shown in
Fig.~\ref{fig:ttbar}.
It has also been suggested~\cite{yung, both1, both2, Buckley, both4} that the
anomalous dijet invariant mass distribution recently
reported by the CDF collaboration in $\ppbar \to \PW + 2$ jets~\cite{CDFbump}
could be evidence for such a boson.

\begin{figure}[htb]
\begin{center}
\includegraphics[width=0.7\linewidth, height=0.2\linewidth]{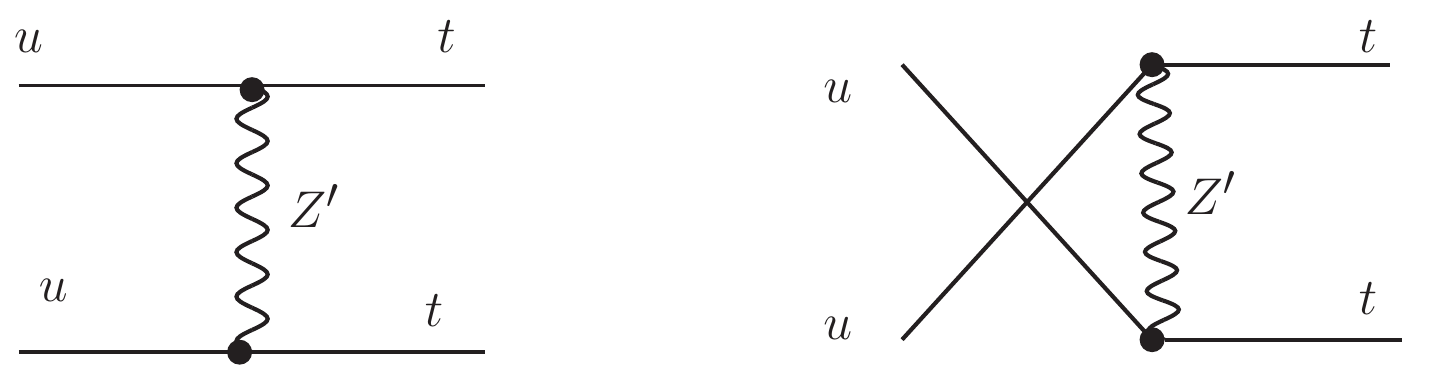}
\includegraphics[width=0.7\linewidth, height=0.25\linewidth]{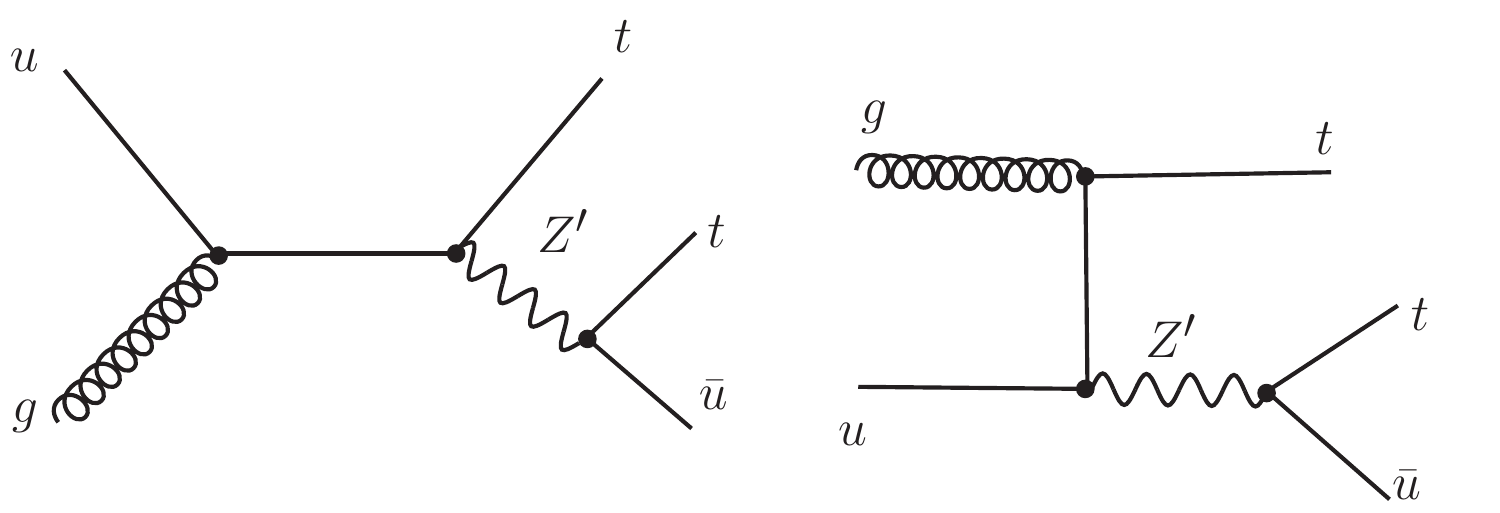}\caption{Diagrams for $\rm tt$ and $\rm ttj$ production in the presence of a $\rm Z'$.
\label{fig:tt}}
\end{center}
\end{figure}

The same type of interaction would also give rise to same-sign
top-quark pair production,
as illustrated in Fig.~\ref{fig:tt}.
In this case, the initial state involves two \PQu quarks.
Because of the large valence quark parton density of the proton,
the \PQt{}\PQt production cross section at the Large Hadron Collider (LHC)
could then be large enough to be observable with a modest amount
of integrated luminosity.  This motivates the search described in this Letter.

For concreteness, we consider the model of Ref.~\cite{berger}.
The relevant \PQu-\PQt-\PZpr interaction term in the Lagrangian is

\begin{equation}
\label{eqn:L_berger}
  \mathcal{L} = g_W \PAQu \gamma^\mu (f_L P_L + f_R P_R)\PQt\PZpr_\mu + \text{h.c.},
\end{equation}

where $g_W$ is the weak coupling strength, and $P_L$
and $P_R$ are the left-handed and right-handed projection
operators.
The left-handed coupling is set to $f_L = 0$, because of
the \PBd-\PABd mixing constraint~\cite{Cao}.
The right-handed coupling $f_R$ and the \PZpr mass ($M_{\PZpr}$)
are the free
parameters of the model.
Within this model there is a narrow range of parameter space
consistent with the Tevatron measurements of $\sigma(\ppbar \to \ttbar)$
and $A_{FB}$, which is not excluded by direct searches for same-sign top-quark pairs~\cite{berger}.

We search for $\pp \to \rm tt$ or $\pp \to \rm ttj$
using a data sample corresponding to an integrated luminosity of 35\pbinv,
collected at a centre-of-mass energy of 7\TeV in 2010 by the
CMS experiment~\cite{CMS} at the LHC.
We concentrate on the final state where both top quarks decay
as $\PQt \to \PW\PQb$ followed by $\PW \to \ell \nu$, where $\ell = \Pe$ or
$\PGm$.  Thus, the final state of interest contains two high
transverse momentum ($p_T$)
isolated positive
leptons (\Pep\ or \PGmp),
two or more jets, and missing transverse energy ($\met$)
from two neutrinos.

The result presented here is based on an already published search for
new physics in events with same-sign isolated dileptons, jets, and
\met~\cite{sspaper}.  In that search we presented event yields and
background expectations for a number of event selections
sensitive
to different possible new physics contributions.  We observed no
excess of events for any of the event selections, and we established
95\% confidence level (CL) upper limits on non-standard model event yields.

We now reinterpret the results of Ref.~\cite{sspaper} in
terms of the model of Eq. (1).  We start from the
``baseline event selection'' defined in \cite{sspaper}.  Briefly,
this selection consisted of two same-sign isolated leptons
of $p_T > 10$ GeV, one of which must have $p_T > 20$ GeV,
at least two jets of $p_T > 30$ GeV and $\met > 20~(30)$ GeV for \Pe{}\PGm
(\PGm{}\PGm\ or \Pe{}\Pe).  Leptons and jets were reconstructed in
the pseudorapidity ranges $|\eta| < 2.4$ and $|\eta| < 2.5$, respectively.
Three events passed these requirements, with an expected background of $3.4 \pm 1.8$.

For the purpose of the tt search, the baseline event selection is
modified by raising the transverse momentum threshold on the
lowest $p_T$ lepton from 10 to 20 GeV, since leptons from
$\PQt\to\PW\PQb$, $\PW\to \ell \nu$
tend to have high $p_T$.  In addition, we demand that both leptons
be positive since we are searching for
$\pp\to\PQt\PQt$ and not $\pp\to \tbartbar$.
These changes are expected to reduce the background by more than
a factor of two.

These requirements select 2 events, while the background expectation
is $0.9 \pm 0.6$ events.
The main background contribution is attributed to $\ttbar$ events with
one lepton from W decay and one ``fake'' lepton,
i.e. a lepton from heavy flavour decay, an electron from
unidentified photon conversion, or a muon from meson decays in flight
and other processes.  This background is estimated in a data-driven way
from studies of the sample of events selected with looser lepton
isolation and identification requirements.
More details on the event selection and the background estimation
can be found in Ref.~\cite{sspaper}.

As a cross-check, we repeat the analysis with the same kinematic
selection, but requiring the two leptons to be negative.
No events in the data satisfy these requirements, in agreement
with the charge-symmetric background expectation
of $0.9 \pm 0.6$ events.

There is no statistically significant excess of tt candidate events
over the background prediction.  We thus proceed to set limits on the
parameters of the Lagrangian of Eq. (1).  The ingredients
for this calculation are the observed (2) and expected ($0.9 \pm 0.6$)
number of events, the cross section for pp$\to$tt(j) as
a function of $f_R$ and $M_{\PZpr}$,
the efficiency of the event selection, the integrated luminosity,
and all associated uncertainties.

We used
the external model interface in the \textsc{MadGraph}~\cite{madgraph}
event generator to calculate at the leading order (LO) the $\PQt\PQt+\PQt\PQt\text{j}$ cross section
as a function of $f_R$ and $M_{\PZpr}$.
In this calculation we used the CTEQ6L~\cite{cteq6l}
parton distribution functions (PDFs),
fixed the top quark mass $M_\text{top}$ to be 172.5 GeV, and set the renormalization
and factorization scales to be $\mu = M_\text{top}$.

{\sc MadGraph} was also used
to generate pp$\to$tt and pp$\to$ttj events
according to the diagrams of Fig.~\ref{fig:tt}.
These
events were then processed by \textsc{Pythia}~\cite{pythia64} for parton
showering, followed by the CMS parametrized event simulation,
and the same chain
of reconstruction and analysis programs used for collision data.
The event selection efficiency, including all relevant branching
ratios, is (0.95 $\pm$ 0.13)\%, independent of \PZpr\ mass.
Note that the branching ratio for tt$\to \ell \nu {\rm b} \ell \nu {\rm b}$
($\ell = {\rm e}, \mu$)
is
$4.54\%$~\cite{bayesian}, but our selection is also sensitive to leptons from
tau decays.
The fractional systematic uncertainty on the event selection efficiencies
was calculated as in Ref.~\cite{sspaper}, and
its components are summarized
in Table~\ref{tab:systSumm}.

\begin{table}[h]
\begin{center}
\caption{\small\label{tab:systSumm}Fractional
systematic uncertainties on the $\pp\to\PQt\PQt(\text{j})$ signal selection efficiency.
ISR/FSR denote initial and final state
radiation.}
\begin{tabular}{lcccc}\hline
Source                                  & \Pe{}\Pe          & \PGm{}\PGm              & \Pe{}\PGm                        & all \\ \hline
Lepton selection                        & 11.8\%                & 10.6\%                & 10.8\%                        & 10.7\% \\
Energy scale                            & 8\%           & 8\%           & 8\%                   & 8\% \\
ISR/FSR and PDF                         & 3\%           & 3\%           & 3\%                   & 3\%   \\
Total without luminosity                & 14.6\%                & 13.6\%                & 13.8  & 13.7\%        \\ \hline
Integrated luminosity                   & 4\%           & 4\%           & 4\%                   & 4\%   \\ \hline
Total & 15\% & 14\% & 14\%  & 14\% \\
\hline
\end{tabular}
\end{center}
\end{table}

\begin{figure}[htb]
\begin{center}
\includegraphics[width=0.7\linewidth]{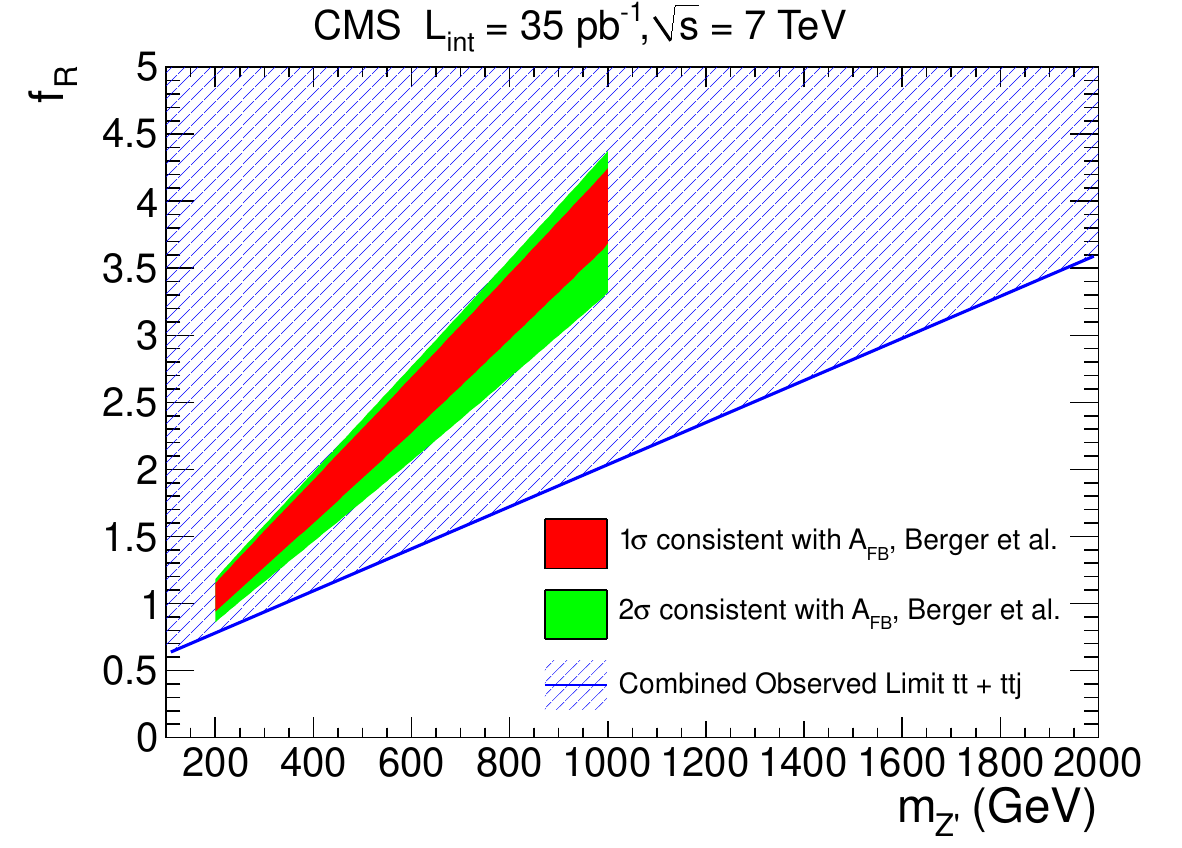}
\caption{\protect The shaded area is the exclusion region at 95\% CL as a function of \PZpr
mass and right handed couplings $f_R$ (see the Lagrangian of Eq. (1)). 
We also show the region of parameter space consistent with
the Tevatron measurements of $A_{FB}$ and $\sigma(\ttbar)$
as inferred in Ref.~\cite{berger}.
\label{fig:bottomline}}
\end{center}
\end{figure}

We compute the upper limits using a Bayesian
method~\cite{bayesian}.  We assume a
flat prior for the signal strength and a log-normal distribution
for the nuisance
parameters.  The 95\% CL upper limit on the number of signal events
is 5.7 using a 14\% uncertainty on the signal efficiency
(Table~\ref{tab:systSumm}).  The expected upper limit is
$4.4^{+1.4}_{-1.3}$ events.
The limit on the cross section is $\sigma(\pp \to \PQt\PQt(\text{j})) < 17.0$ pb at 95\% CL.
As a cross-check, we also calculate the upper limit on the number
of events using the hybrid frequentist-Bayesian CL$_s$ approach~\cite{CLS}.
With this method we find an upper limit of 5.6 events, which demonstrates
the insensitivity of our results to the details of the
statistical analysis.

Using the LO cross section as computed
in \textsc{MadGraph}, we turn the
limit on the number of
events into an exclusion region in the $f_R$-$M_{\PZpr}$ plane as
shown in Fig.~\ref{fig:bottomline}.  The region of parameter
space consistent with the Tevatron $A_{FB}$~\cite{berger} is disfavoured
by this analysis.

The Lagrangian of Eq. (1) has two parameters, $M_{\PZpr}$ and $f_R$.
At very large values of $M_{\PZpr}$ this lagrangian becomes equivalent
to $\mathcal{L} = -\frac{1}{2}\frac{C_{RR}}{\Lambda^2}
 [\PAQu_R \gamma^\mu \PQt_R][\PAQu_R \gamma_{\mu} \PQt_R] + \text{h.c.}$~\cite{Jung:2009pi, cdfth2},
with $\frac{C_{RR}}{\Lambda^2} = \frac{2 g_W^2 f_R^2}{M_{\PZpr}^2}$.
Using the $f_R$ limit calculated at $M_{\PZpr} = 2$ TeV, 
which is the highest value of the $\PZpr$-mass considered in
our analysis, we set a limit 
$\frac{C_{RR}}{\Lambda^2} < 2.7$ TeV$^{-2}$ at
95\% confidence level.
This bound is more stringent than that recently reported
by CDF: $\frac{C_{RR}}{\Lambda^2} < 3.7$ TeV$^{-2}$~\cite{cdflimit}.

In summary, we have established a limit on tt production in pp
collisions at $\sqrt{s} = 7$ TeV, based on a search for
same-sign dileptons.  Our
bound can
be used to test models of new physics with massive \PZpr
bosons that have been proposed to explain the
Tevatron measurements of the $\ppbar \to \ttbar$
forward-backward asymmetry.
Our result disfavours this FCNC interpretation.

We thank Johan Alwall, Ed Berger, Qing-Hong Cao, Chuan-Ren Chen, Chong-Sheng Li
and Hao Zhang for discussions and help in implementing the $Z'$ model in
\textsc{MadGraph/MadEvent}.
We wish to congratulate our colleagues in the CERN accelerator departments for the excellent performance of the LHC machine. We thank the technical and administrative staff at CERN and other CMS institutes, and acknowledge support from: FMSR (Austria); FNRS and FWO (Belgium); CNPq, CAPES, FAPERJ, and FAPESP (Brazil); MES (Bulgaria); CERN; CAS, MoST, and NSFC (China); COLCIENCIAS (Colombia); MSES (Croatia); RPF (Cyprus); Academy of Sciences and NICPB (Estonia); Academy of Finland, ME, and HIP (Finland); CEA and CNRS/IN2P3 (France); BMBF, DFG, and HGF (Germany); GSRT (Greece); OTKA and NKTH (Hungary); DAE and DST (India); IPM (Iran); SFI (Ireland); INFN (Italy); NRF and WCU (Korea); LAS (Lithuania); CINVESTAV, CONACYT, SEP, and UASLP-FAI (Mexico); PAEC (Pakistan); SCSR (Poland); FCT (Portugal); JINR (Armenia, Belarus, Georgia, Ukraine, Uzbekistan); MST and MAE (Russia); MSTD (Serbia); MICINN and CPAN (Spain); Swiss Funding Agencies (Switzerland); NSC (Taipei); TUBITAK and TAEK (Turkey); STFC (United Kingdom); DOE and NSF (USA).

\bibliography{auto_generated}   

\cleardoublepage \appendix\section{The CMS Collaboration \label{app:collab}}\begin{sloppypar}\hyphenpenalty=5000\widowpenalty=500\clubpenalty=5000\textbf{Yerevan Physics Institute,  Yerevan,  Armenia}\\*[0pt]
S.~Chatrchyan, V.~Khachatryan, A.M.~Sirunyan, A.~Tumasyan
\vskip\cmsinstskip
\textbf{Institut f\"{u}r Hochenergiephysik der OeAW,  Wien,  Austria}\\*[0pt]
W.~Adam, T.~Bergauer, M.~Dragicevic, J.~Er\"{o}, C.~Fabjan, M.~Friedl, R.~Fr\"{u}hwirth, V.M.~Ghete, J.~Hammer\cmsAuthorMark{1}, S.~H\"{a}nsel, M.~Hoch, N.~H\"{o}rmann, J.~Hrubec, M.~Jeitler, W.~Kiesenhofer, M.~Krammer, D.~Liko, I.~Mikulec, M.~Pernicka, B.~Rahbaran, H.~Rohringer, R.~Sch\"{o}fbeck, J.~Strauss, A.~Taurok, F.~Teischinger, P.~Wagner, W.~Waltenberger, G.~Walzel, E.~Widl, C.-E.~Wulz
% ...
\vskip\cmsinstskip
% ...
1:~~Also at CERN, European Organization for Nuclear Research, Geneva, Switzerland\\
% ...

\end{sloppypar}
\end{document}